\documentclass[desactivate]{aa}
\usepackage[varg]{txfonts}

\usepackage{natbib}
\usepackage[T1]{fontenc}
\usepackage{geometry}
\usepackage{graphicx}
\usepackage{tabto}
\usepackage{float}
\usepackage{caption}
\usepackage{subcaption}
\usepackage{hhline}
\usepackage{setspace}
\bibpunct{(}{)}{;}{a}{}{,}

\newcommand{\mum}{$\mathrm{\mu m}$}
\newcommand{\CII}{[C{\sc ii}]\,158$\,\mathrm{\mu m}$}
\newcommand{\OI}{[O{\sc i}]\,63$\,\mathrm{\mu m}$}
\newcommand{\OIb}{[O{\sc i}]\,145$\,\mathrm{\mu m}$}
\newcommand{\OIV}{[O{\sc iv}]\,26$\,\mathrm{\mu m}$}
\newcommand{\NeV}{[Ne{\sc v}]\,$14\,\mathrm{\mu m}$}
\newcommand{\hii}{H{\sc ii}}
\newcommand{\Zsol}{$\mathrm{Z_\odot}$}
\newcommand\modif[1]{#1}
\newcommand\modiftwo[1]{#1}

\def\aap{A\&A}

\def\apjl{ApJL}
\def\apjs{ApJS}

\usepackage{orcidlink}

\title{Probing the heating of the neutral atomic interstellar medium in the Dwarf Galaxy Survey through infrared cooling lines}
\titlerunning{Probing the heating of the neutral ISM in the Dwarf Galaxy Survey}
\author{M. Varese \inst{\ref{inst1}} \orcidlink{0009-0004-6699-8341}\and
        V. Lebouteiller \inst{\ref{inst1}} \orcidlink{0000-0002-7716-6223}\and
        L. Ramambason \inst{\ref{inst2}} \orcidlink{0000-0002-9190-9986} \and
        F. Galliano \inst{\ref{inst1}} \orcidlink{0000-0002-9190-9986} \and
        C.~T. Richardson \inst{\ref{inst3}} \orcidlink{0000-0002-3703-0719} \and
        S.~C. Madden \inst{\ref{inst1}} \orcidlink{0000-0003-3229-2899}}

\institute{
Université Paris-Cité, Université Paris-Saclay, CEA, CNRS, AIM,
91191 Gif-sur-Yvette, France \label{inst1}
\and
Institut fur Theoretische Astrophysik, Zentrum für Astronomie,
Universität Heidelberg, Albert-Ueberle-Str. 2, D-69120 Heidelberg \label{inst2}
\and
Physics Department, Elon University, 100 Campus Drive, Elon, NC 27244, USA  \label{inst3}
}
\date{Received {} /
      Accepted {}  }       

\abstract  {Star formation in galaxies is regulated by dynamical and thermal processes. In the Milky Way and star-forming galaxies with similar metallicity, the photoelectric effect on small dust grains usually dominates the heating of the neutral atomic gas, which constitutes the main star-forming gas reservoir. In more metal-poor galaxies, the lower dust-to-gas mass ratio together with the higher occurrence and luminosity of X-ray sources suggest that other heating mechanisms may be at play.}
{We aim to determine the contribution of the photoelectric effect, photoionization by UV and X-ray photons as well as ionization by cosmic rays to the total heating of the neutral gas in a sample of 37 low-metallicity galaxies. In particular, we wish to assess whether X-ray sources can be a significant source of heating. \modif{We also} attempt to recover the intrinsic X-ray fluxes and compare with observations when available.} 
{We use the statistical code MULTIGRIS together with a photoionization grid of Cloudy models propagating radiation from stellar clusters and potential X-ray sources to the ionized and neutral gas. \modif{This grid includes physical parameters such as metallicity, gas density, ionization parameter, and radiative source properties.} We describe a galaxy as a combination of many 1D components linked by few physical hyperparameters. Infrared cooling lines are used as constraints to evaluate the most likely combinations and parameters.} 
{We constrain the heating fractions for the main mechanisms for the first time in a low-metallicity galaxy sample. We show that for the higher metallicity galaxies, the photoelectric effect dominates the neutral gas heating. At metallicities below $1/8$ the Milky Way value, cosmic rays and photoionization can become predominant. \modif{We compute an observational proxy for the photoelectric effect heating efficiency on polycyclic aromatic hydrocarbons (PAHs) using the total cooling traced by \CII + \OI. We show that this proxy can match theoretical expectations when accounting for the fraction of the heating due to the photoelectric effect according to our models.} Finally, we show that it is possible to predict reasonably well the X-ray fluxes in the $0.3-8$\,keV band from the gas cooling lines for most of the galaxies observed in this band. With the current grid and assumptions, it remains difficult to conclude on the exact heating fraction due to cosmic rays, but we speculate that heating from X-ray sources is more important. } 
{As expected from the low abundance of dust and PAHs in metal-poor galaxies, other heating mechanisms than the photoeletric effect heating must be accounted for. Bright X-ray sources may deposit their energy on large scales in such transparent, dust-poor interstellar medium, and thus represent promising avenues to understand the physical properties of the main star-forming gas reservoir in galaxies. The modelling strategy adopted here makes it possible to recover the global intrinsic radiation field properties when X-ray observations are unavailable, for instance in early universe galaxies.}

\begin{document}
\maketitle

\section{Introduction}

While star formation mainly occurs in dense, cold neutral \modif{molecular} gas, the main gas reservoir over long timescales is \modif{warmer (few $10^{2-3}$\,K)} and atomic. For typical Gyr timescales, star formation is regulated by thermodynamical processes, that control the transformation of warm neutral \modif{atomic} gas into cold \modif{neutral} molecular gas (e.g., \citealt{chevance_life_2022}). Heating of the ionized gas of \hii\ regions is known to be dominated by UV photoionization of H and He, but the neutral gas heating is comparatively less well known. It is due to many potential mechanisms (photoelectric effect, cosmic ray (CR) ionization, X-ray photoionization, shocks, etc - see \citealt{dalgarno_heating_1972,girichidis_physical_2020}) that are difficult to disentangle, many of which are linked to the production of ionizing photons or energetic events that inform us on important galactic parameters such as the star formation rate (SFR). Since the heating mechanisms are difficult to probe directly, we can observe fine-structure cooling lines instead, which allow us to infer several diagnostics, assuming thermal equilibrium. 

At solar metallicity (Z$_\odot$), neutral atomic gas heating is generally dominated by the photoelectric effect on dust grains (e.g., \citealt{weingartner_photoelectric_2001}). The majority of the photoelectric effect heating is due to small grains and polycyclic aromatic hydrocarbons (PAH; \citealt{berne_contribution_2022}). Since the dust-to-gas mass ratio (DGR) decreases with metallicity \citep{galliano_nearby_2021}, as does the PAH abundance (PAH emission is not detected under 1/10\,Z$_\odot$, \citealt{remy-ruyer_linking_2015}), a relatively lower photoelectric effect heating rate is expected at lower metallicities even if small, non-PAH grains may still contribute to the heating. It could be compensated by mechanisms such as cosmic ray heating, photoionization by X-rays, or shocks. The theoretical study from \citet{bialy_thermal_2019} indeed shows that CR heating can become prominent at low metallicity. \modiftwo{Similar results are found by Brugaletta at al. (arXiv:2410.19087) in magneto-hydrodynamic simulations including CR propagation in low-metallicity galaxies.} However an observationally-driven confirmation is still needed. Moreover, the increasing luminosity and number of luminous X-ray sources in metal-poor galaxies \citep{brorby_x-ray_2014, lehmer_metallicity_2021, cann_exploring_2024} suggest that photoionization by X-rays can also be an important source of heating at low metallicity. This has been explored by \citet{pequignot_heating_2008} and \citet{lebouteiller_neutral_2017} in the extremely low metallicity star-forming dwarf galaxy I\,Zw\,18 ($\approx1/35$\,Z$_\odot$), showing that X-ray heating should dominate over the photoelectric effect and proposing that CR ionization may not be important. \modif{The relative contribution of photoelectric effect, cosmic rays, and photoionization by X-rays to the heating of the neutral gas has been investigated in the Milky Way by \citet{wolfire_neutral_2003}. However, there is no theoretical study to our knowledge that includes the contribution of X-ray photoionization to the heating at very-low metallicity.}

From an observational perspective, it is particularly difficult to distinguish CR heating and X-ray heating, since both mechanisms effectively lead to the ionization of the same species. However, since CRs and X-rays penetrate at different depths within a cloud (depending on the X-ray photon energy), several studies attempt to differentiate them by probing the chemical network and its resulting tracers especially at high extinction (e.g., ionized molecules; \citealt{wolfire_photodissociation_2022}). CO emission has also been used to probe X-ray dominated regions (e.g., \citealt{mingozzi_co_2018, vallini_impact_2019}) and to distinguish them from cosmic ray dominated regions \citep{meijerink_irradiated_2006} even though a strong conclusion is difficult to reach (e.g., \citealt{rigopoulou_herschel-spire_2013, rosenberg_molecular_2014}). The molecular gas is unfortunately and notoriously difficult to probe in extremely metal-poor galaxies (e.g., \citealt{cormier_molecular_2014, hunter_interstellar_2024}) and we focus here instead on the neutral atomic gas (with some contribution from the so-called CO-dark molecular gas, \citealt{wolfire_dark_2010, madden_tracing_2020}) which is mainly cooling down through IR cooling lines \CII\ and \OI\ \modif{\citep{tielens_photodissociation_1985-1, tielens_photodissociation_1985,wolfire_neutral_2003}}.

While X-ray sources are expected to play an important role in low metallicity dwarf galaxies \citep{lebouteiller_neutral_2017}, their exact nature is generally unknown. Ultraluminous X-ray sources (ULXs, with $L_{\rm X}>10^{39}\,\mathrm{erg\,s^{-1}}$; \modiftwo{see review in \citep{kaaret_ultraluminous_2017}}) have been detected in several dwarf galaxies, but the intrinsic shape of their spectral energy distribution (SED) remains unknown and seems to vary from one object to another \citep{ott_chandra_2005, ott_chandra_2005-1}. However, the ULX SED shape, and therefore the associated hints on the ULX nature, may be inferred indirectly from high ionization potential lines \modif{of the ISM irradiated by these compact sources} \citep{garofali_modeling_2024}.

In this paper, we wish to determine the contribution of the photoelectric effect, cosmic ray, and X-ray photoionization to the total heating of neutral atomic ISM in the Dwarf Galaxy Survey (DGS; \citealt{madden_overview_2013}), a sample of dwarf galaxies covering a wide range of metallicities (from $1/35$\,Z$_\odot$ to $1/2$\,Z$_\odot$). To this aim, we propose using adequate radiative transfer models in order to link heating and cooling processes, using infrared (IR) cooling lines as our main constraints.
As a secondary aim, we wish to recover the X-ray fluxes of galaxies in the DGS where ULXs have been observed, using in particular \OIV\ and \NeV\ as main constraints. Given their high ionization potentials, these ionized ISM lines are appropriate to probe the X-ray sources. \modiftwo{Such high ionization potential lines have already been used to probe ULX properties in \citet{kaaret_photoionized_2009}}. Nevertheless, X-rays may also penetrate in the neutral medium and heat it, implying that lines such as \CII\ or \OI\ can provide additional constraints on the X-ray fluxes.

We describe in Section\,\ref{sec:observations} the DGS sample and the corresponding observations in IR and X-rays. The modelling strategy is presented in Section\,\ref{sec:method}, along with the statistical framework MULTIGRIS used in this work. The main results are presented in Section\,\ref{sec:results}. Potential improvements concerning cosmic rays and X-ray sources are discussed in Section\,\ref{sec:discussion}.

\section{Observations}

\label{sec:observations}

For this study, we use IR and X-ray observations of a sample of low-metallicity dwarf galaxies. IR lines are used to trace the cooling of the neutral and ionized ISM, allowing us to infer physical conditions and potential heating mechanisms. X-ray observations are used to probe the presence of luminous X-ray binaries that produce significant ionization of the ISM. While we do not use the X-ray observations as direct constraints, we compare them in the following with the inferred X-ray source properties. 

\subsection{Sample}\label{subsec:DGS}

We use the sample from \citet{ramambason_inferring_2022}, which is a extracted from the Dwarf Galaxy Survey (DGS; \citealt{madden_overview_2013}). Our sample contains 37 local star-forming dwarf galaxies ($0.7-191$\,Mpc) with a metallicity spanning $1/35$\,\Zsol\ to $1/2$\,\Zsol, making it suitable for studying the variation of the heating mechanisms as a function of metallicity. We restrict the sample to spatially-unresolved galaxies in order to include only objects with a coherent line set probing the full galaxy. This is also meant to match the method used on Section\,\ref{sec:method} which models entire galaxies. Previous studies on the DGS, based on SED fitting, provide us with various parameters such as the mass of neutral and molecular hydrogen or the DGR \citep{remy-ruyer_gas--dust_2014, remy-ruyer_linking_2015, galliano_nearby_2021}.

\subsection{Infrared spectroscopic observations}\label{subsec:IRobs}

DGS galaxies have been observed with \textit{Spitzer} (mid-IR) and \textit{Herschel} (far-IR), providing spectroscopic and photometric observations (emission lines and total IR luminosity; \citealt{cormier_herschel_2015, cormier_herschel_2019}). Cooling lines in the neutral medium such as \CII\ and \OI\ are frequently detected (in 37 and $27$ galaxies, respectively). We also have access to tracers of the ionized gas (e.g., [NeIII], [NeII]) including from highly ionized species such as O$^{\rm 3+}$ ([O\,{\sc iii}] \modiftwo{detected} in $17$ galaxies) or Ne$^{\rm 4+}$ (with [Ne\,{\sc v}] upper limits available for $31$ galaxies).

For $27$ galaxies in our sample, PAH emission is measured from \textit{Spitzer}/IRS spectra (from the CASSISjuice database, \citealt{lebouteiller_cassis_2011, lebouteiller_cassis_2015, lebouteiller_cassisjuice_2023}) with the MILES spectral decomposition software, which uses the model from \citet{galliano_variations_2008} together with the band profiles given in Table\,III.4 of \citet{galliano_nearby_2022} to decompose the spectra into dust continuum, PAH emission bands, emission lines, and silicate absorption bands. This gives us access to the total PAH emission between $5$\,\mum~and $14$\,\mum. For $18$ sources with low signal-to-noise ratio and/or low PAH abundance, this method provides an upper limit on the PAH emission.

\subsection{X-ray observations}
\label{subsec:Xobs}

\modif{In order to compare the X-ray luminosities predicted by our models to actual observations, we searched the literature for observations of X-ray sources in the galaxies of our sample. Observations in the $0.3-10$\,keV range from \textit{Chandra}, XMM-\textit{Newton}, and eROSITA are available for $15$ galaxies \citep{thuan_chandra_2004, ott_chandra_2005, rosa_gonzalez_evolution_2009, kaaret_x-rays_2011, oti-floranes_multiwavelength_2012, kaaret_state_2013,prestwich_ultra-luminous_2013, brorby_x-ray_2014, prestwich_ultra-luminous_2015, gross_resolving_2021, bykov_srgerosita_2023}. Due to sensitivity, most of the observations are dominated by ULXs, except for two diffuse emission regions in He2-10 and NGC1569. All these studies use a given physical model prescription for the X-ray spectrum and line-of-sight absorption, which, together, enable the recovery of the intrinsic X-ray spectrum, from which luminosities are computed.}

\modif{The X-ray emission of accreting black holes is thought to arise from two main components: a compton corona, whose emission is described by a power-law, and an accretion disk, modelled by a multicolor blackbody (see review in \citealt{remillard_x-ray_2006}). Since the compton emission is relatively more easily detected because less absorbed, all of the above studies compute the X-ray luminosities using a power-law. A blackbody (potentially multicolor) is also used in \citet{thuan_chandra_2004}, \citet{ott_chandra_2005-1}, and \citet{kaaret_state_2013} as a comparison.}

\modif{We compiled $70$ luminosities in $8$ different energy bands within $0.3-10$\,keV, some of them arising from the same observation processed with different physical models. Considering the relatively small differences in the luminosities obtained with different models (typically less than a factor of $2$), it would be beyond the scope of the present study to homogenize all observations by forcing the same physical model and we average instead the luminosities computed from different models. However, we have chosen not to combine the luminosities coming from different studies altogether, to prevent biases linked to the instrument or data reduction. This effectively leads to multiple observations for few galaxies such as IZw18 (see Table\,\ref{tab:Xrayobs}). The associated error bar accounts for the error on each prescription and the scatter of the different prescriptions.}

\modif{In some galaxies, multiple sources are observed. When this happens, we sum their luminosities to get the total X-ray luminosity of the galaxy. It is likely that dwarf galaxies host a collection of X-ray sources, of which ULXs represent the brightest tail. However ULXs are believed to account for the majority of the total X-ray luminosity  \citep{garofali_modeling_2024}. The final luminosities obtained with this method can be found in Table\,\ref{tab:Xrayobs}.}

\begin{table*} 
\caption{X-ray observations of DGS galaxies.}
\label{tab:Xrayobs}
\label{tab:} 
   \centering 
   \begin{tabular}{l c c c c c c c c c c c c c c c c } 
\hline\hline 
      Galaxy & $L_{0.3-2\,{\rm keV}}$ & $L_{0.3-8\,{\rm keV}}$ & $L_{0.5-8\,{\rm keV}}$ & $L_{2.0-10\,{\rm keV}}$ & ref \\ \hline 
      HS0822+3542 &  &  &  & $<1.6\times 10^{38}$ & (6) \\ 
      HS0822+3542 &  & <$2.0\times 10^{38}$ &   &  & (8) \\ 
      HS1442+4250 &  &   & $2.2\times 10^{38}$ &  & (1) \\ 
      HS1442+4250 &  &   &  & $<1.2\times 10^{38}$ & (6) \\ 
      HS1442+4250 &  & $<1.3\times 10^{38}$ &  &  & (8) \\ 
      Haro11 & $6.8\times 10^{40}\ ^{+9.0\times 10^{39}}_{-8.0\times 10^{39}}$ & $1.6\times 10^{41}\pm 1.0\times 10^{40}$ &  &  & (4) \\ 
      Haro11 &  & $1.5\times 10^{41}\ ^{+1.4\times 10^{41}}_{-2.3\times 10^{40}}$ &  &  & (9) \\ 
      Haro2 & $2.2\times 10^{40}\ ^{+6.7\times 10^{39}}_{-5.7\times 10^{39}}$ & $4.2\times 10^{40}\pm 8.2\times 10^{39}$ & $3.6\times 10^{40}\pm 7.1\times 10^{39}$ &  & (2) \\ 
      Haro2 &  & $1.8\times 10^{39}\ ^{+2.2\times 10^{38}}_{-2.0\times 10^{38}}$ &  &  & (3) \\ 
      He2-10 &  & $8.2\times 10^{40}\pm 2.3\times 10^{40}$ &  &  & (7) \\ 
      IZw18 &  &  & $3.8\times 10^{39}$ &  & (1) \\ 
      IZw18 & $2.7\times 10^{39}\ ^{+7.0\times 10^{38}}_{-6.1\times 10^{38}}$ & $5.0\times 10^{39}\pm 8.7\times 10^{38}$ & $4.3\times 10^{39}\pm 7.5\times 10^{38}$ &  & (2) \\ 
      IZw18 &  &  &  & $1.3\times 10^{39}$ & (6) \\ 
      IZw18 &  & $2.7\times 10^{39}\ ^{+1.1\times 10^{40}}_{-2.7\times 10^{39}}$ &  &  & (7) \\ 
      IZw18 &  & $4.9\times 10^{39}\pm 2.5\times 10^{38}$ &  &  & (8) \\ 
      Mrk209 &  &  &  & $<6.6\times 10^{37}$ & (6) \\ 
      Mrk930 &  &  &  & $1.2\times 10^{40}\pm 4.0\times 10^{39}$ & (10) \\ 
      NGC1569 &  & $8.3\times 10^{38}\pm 1.2\times 10^{38}$ &  &  & (7) \\ 
      NGC5253 &  & $1.4\times 10^{39}\pm 4.2\times 10^{38}$ &  &  & (7) \\ 
      SBS0335-052 &  & $1.9\times 10^{39}\pm 4.2\times 10^{38}$ &  &  & (8) \\ 
      SBS1415+437 &  & $<1.3\times 10^{38}$ &  &  & (8) \\ 
      UGC4483 &  &  &  & $<2.2\times 10^{37}$ & (6) \\ 
      UM461 &  &  &  & $<3.3\times 10^{37}$ & (6) \\ 
      VIIZw403 &  &  & $4.0\times 10^{38}$ &  & (1) \\ 
      VIIZw403 &  &  &  & $3.5\times 10^{38}$ & (6) \\ 
      VIIZw403 &  & $3.3\times 10^{38}\ ^{+4.6\times 10^{38}}_{-3.0\times 10^{38}}$ &  &  & (7) \\ 
   \hline  
   \end{tabular} 
\tablefoot{The luminosities are given in erg\,s$^{-1}$. When different models are available, their luminosities are averaged as explained in Sect.\ref{subsec:Xobs}.}
\tablebib{(1) \citet{brorby_x-ray_2014}; (2) \citet{bykov_srgerosita_2023}; (3) \citet{grier_discovery_2011}; (4) \citet{gross_resolving_2021}; (5) \citet{kaaret_state_2013}; (6) \citet{kaaret_x-rays_2011}; (7) \citet{ott_chandra_2005-1}; (8) \citet{prestwich_ultra-luminous_2013}; (9) \citet{prestwich_ultra-luminous_2015}; (10) \citet{rosa_gonzalez_evolution_2009}; (11) \citet{thuan_chandra_2004}.} 
\end{table*}

\section{Method}\label{sec:method}

\subsection{Strategy}\label{subsec:combination}

We use 1D photoionization models as basic components to represent our galaxies. These models contain a central source \modiftwo{(which is a star cluster, with or without a compact object)} illuminating a slab or shell with a given density, metallicity, and depth. In order to describe a full galaxy with such models, we combine them through a statistical distribution. This method is meant both to use a physically motivated distribution and to keep a reasonable number of parameters ($15$ in the chosen ``architecture'' -- described in Sect.\,\ref{subsec:inference}).

With this method, the emission of an observable $L$ is defined as 
\begin{equation}L_{\rm tot} = \sum_{\bf p_{min}}^{\bf p_{max}} \prod_{i=0}^n \Phi({p_i}) I({\bf p}) \Delta({p_i})\end{equation}
with $\Phi(p_i)$ the weight associated with the parameter $p_i$ according to some statistical distribution, $I({\bf p})$ the observable for a given parameter set, and $\Delta({p_i})$ the grid parameter interval. The weight $\Phi(p_i)$ is determined by the statistical distribution used. For instance, a power-law distribution will be defined as:
\begin{equation}
    \Phi(p) = \left\{
    \begin{array}{ll}
        10^{\alpha_p} & \mbox{if } p \in [p_{\rm min}, p_{\rm max}] \\
        0 & \mbox{}
    \end{array}
  \right\},\label{eq:pl}\end{equation}
while a broken power-law would be defined as:
\begin{equation}
\Phi(p) = \left\{
    \begin{array}{ll}
        10^{\alpha_{p1}} & \mbox{if } p \in [p_{\rm min}, p_{\rm 0}] \\
        10^{\alpha_{p2}} & \mbox{if } p \in [p_{\rm 0}, p_{\rm max}] \\
        0 & \mbox{}
    \end{array}
  \right\},\label{eq:bpl}\end{equation}
and a parameter described by a single value as:
\begin{equation}
\Phi(p) = \delta_p = \left\{
    \begin{array}{ll}
        1 & \mbox{if } p = p_{\rm val} \\
        0 & \mbox{}
    \end{array}
  \right\}.\label{eq:sp}\end{equation}

A detailed description of this method is provided in \citet{lebouteiller_topological_2022}, and is used in \citet{ramambason_modeling_2024} and Lebouteiller et al. (in prep.).

\subsection{SFGX model database}
\label{subsec:sfgx}

\begin{figure}
        \centering
        \includegraphics[width=.5\textwidth]{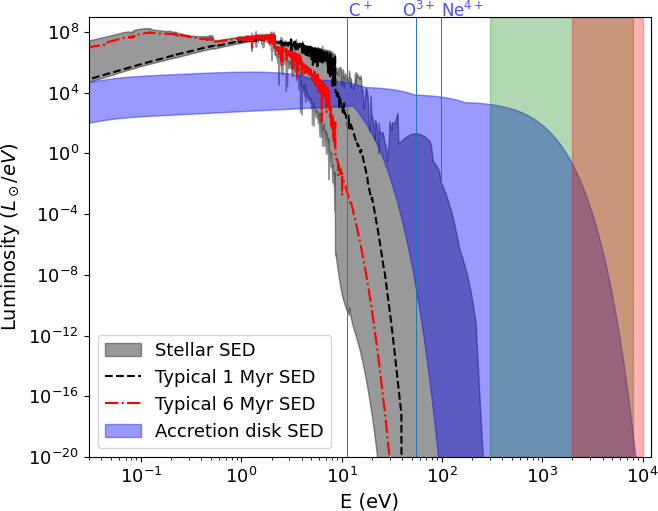}
        \caption{\modif{Stellar SEDs with instantaneous burst age $1$\,Myr (black dashed line) and $6$\,Myr (red dashed line). The grey area represents the spread from the age-metallicity parameter space. The blue area shows the variation of an accretion disk SED when $T_{\rm X}$ is varied (see text)}. The green and red shaded vertical stripes represent the $0.3-8$\,keV and $2-10$\,keV bands.}
        \label{fig:SED}
\end{figure}

In this study, we use the SFGX grid of 1D models (Star Forming Galaxies with X-ray sources; \citealt{ramambason_inferring_2022}), calculated with the code Cloudy 17.02 \citep{ferland_2017_2017} and the stellar population BPASS (version 2.1; \citealt{eldridge_binary_2017}). The latter uses variable stellar age and metallicity, with the stellar metallicity being equal to the ISM metallicity. Physical properties of the ISM such as the metallicity or density are varied. The DGR \modif{varies with metallicity following} the prescription of \citet{galliano_nearby_2021}. SFGX also includes a potential X-ray source whose emission is modeled as a multicolor blackbody, representing an accretion disk \citep{mitsuda_energy_1984}. The inner temperature $T_{\rm X}$ of the accretion disk is varied from $10^5$~K to $10^7$~K, while the outer temperature $T_{\rm out}$ is fixed to $10^3$~K. The luminosity $L_{\rm X}$ is given as a fraction of the bolometric luminosity, spanning from $0$ to $0.1$. The X-ray and stellar SEDs used in SFGX are shown in Figure\,\ref{fig:SED}. A complete description of the grid is provided in \citet{ramambason_inferring_2022}. A synthetic version with the parameter ranges is presented in Table\,\ref{tab:SFGX_sum}.

\begin{table}
\caption{Summary of SFGX primary parameters.}
\label{tab:SFGX_sum}
\centering
\begin{tabular}{l l l}
    \hline\hline
    Parameter & Range & Distribution\\
    \hline
     $\log L$ & [7, 9] L$_\odot$ & Single value\\
     Age of burst & [1, 10] Myr & Single value\\
     $\log n_H$ & [0, 4] $\mathrm{cm^{-3}}$ & Power-law\\
     $\log U^{(1)}$ & [-4, 0] & Power-law\\
     $\log Z$ & [-2.19, 0.111] Z$_\odot$ & Single value\\
     $\log L_{\rm X}$ & $-\infty$, [-3, -1] L$_{\rm bol}$ & Single value $^{(3)}$ \\
     $\log T_{\rm X}$ & [5, 7] K & Single value $^{(3)}$\\
     cut$^{(2)}$ & [0, 4] & Broken power-law\\
     \hline
\end{tabular}
\tablefoot{$^{(1)}$ Ionization parameter, controlled by varying the inner radius of the cloud. $^{(2)}$ Parameter controlling the depth of the cloud (see Sect.\,\ref{subsec:sfgx}). $^{(3)}$ Indicates parameters where linear interpolation is performed between nearest neighbours in the grid.}
\end{table}

SFGX also allows the depth of the model to be a free parameter to describe matter-bounded clouds. Since running a full model is time consuming, one model is run for each set of primary parameters, with the stopping criterion being either the visual extinction ($A_V=10$) or the gas temperature ($T_e = 10$\,K), whichever happens first. These models are then truncated at different depths, creating several sub-models. These sub-models are defined by a parameter named "cut" in SFGX, that determines the depth at which each model has been truncated. In order to make the sub-models comparable for different metallicities, the cut parameter is defined as follows: cut=0: illuminated face of the cloud, $A_V=0$. cut=1: $\mathrm{H^+}$-$\mathrm{H^0}$ transition. cut=2: $\mathrm{H^0}$-H$_2$ transition. cut=3: C-CO transition. cut=4: end of model ($A_V=10$ or $T_e=10$\,K). We then define non-integer cuts ($0.25$ step) by interpolating linearly on $A_V$, resulting in $17$ sub-models per model actually ran. One should keep in mind that Cloudy models cropped to a given cut are not strictly equivalent to models specifically designed to stop at this cut, in particular due to optical depth effects, \modiftwo{but we have verified that these effects are not significant. To this purpose, we ran a set of 1D Cloudy models stopped at different depths to compare them to the cropped models, with the physical parameters being choosen to match the average conditions in the DGS galaxies. We found a difference in line emission below 30\% for \CII\ and \OI, caused by a drop of temperature at the outer edge of the cloud for the non-cropped models. This drop in temperature is likely caused by the absence of radiation beyond the depth at which the model is stopped, while in reality at least the diffuse interstellar radiation field will always illuminate the back of the cloud. Overall, we conclude that the cropped models, which do not show this temperature drop, provide a good enough approximation for line emission at any given depth.}

\begin{figure}
    \centering
    \includegraphics[width=.5\textwidth]{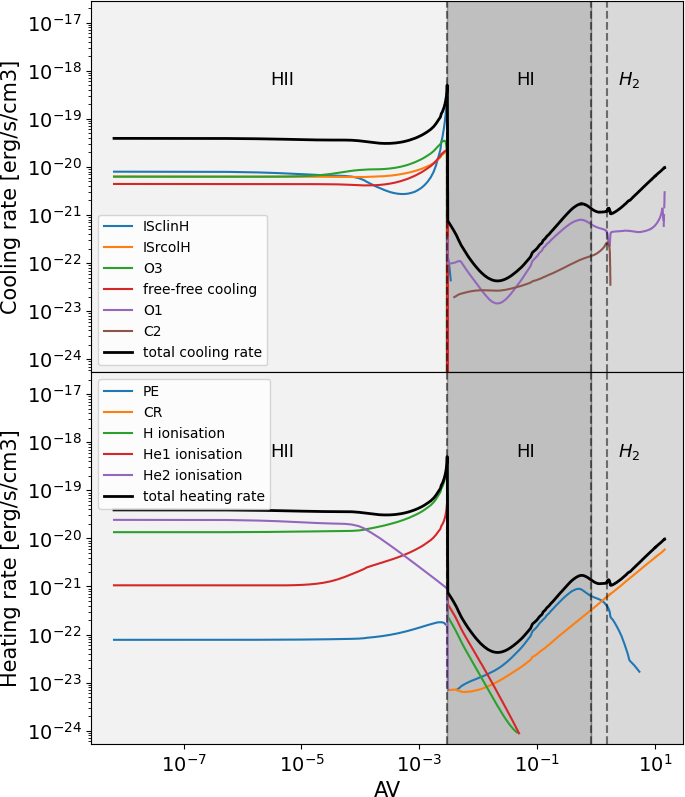}
    \caption{Cooling and heating rate vs. depth for the main mechanisms in a typical 1D model (stellar age: $6$\,Myr, $L_* = 10^9\,L_\odot$, $n=2$, $U=-2$, $Z=0.1\,Z_\odot$, $L_{\rm X} = 10^7\,L_\odot$, $T_{\rm X}=10^6\,\mathrm{K}$). The dashed vertical lines correspond to the cut values 1, 2, and 3. Acronyms: "IScclinH": hydrogen recombination; "ISrcolH": net free-bound cooling minus free-free heating; "PE": photoelectric effect heating; "CR": cosmic ray heating.}
    \label{fig:heatcool1D}
\end{figure}

Cloudy models give us access to the heating rates, cooling rates, and intensity of emission lines at each depth within the cloud (see Fig.\,\ref{fig:heatcool1D}). In SFGX, these values are volume-integrated. Since we wish to analyze the heating mechanisms as a function of depth, we compute the heating and cooling rates integrated between two successive cuts. The same calculation is performed for the emission lines, allowing us to evaluate the fraction of the line emission corresponding to any given depth.

A strong hypothesis in SFGX is that the cosmic ray ionization rate is fixed at $3$ times the Galactic value (i.e., $3\times2\times 10^{-16}\, \mathrm{s}^{-1}$; \citealt{indriolo_documentclassaastex_2007}). This choice was initially motivated by the elevated \OIb/\OI\ ratio in the DGS, that cannot be explained without an enhanced CR ionization rate or a high X-ray luminosity (see \citealt{cormier_herschel_2019} for details). Given that the latter was given a relatively low value, this likely led to a higher CR ionization rate than actually needed.

\subsection{Parameter estimation}\label{subsec:inference}

For each galaxy (i.e., each set of IR observations and the associated error bars), the best combination of 1D models is inferred using the Bayesian code MULTIGRIS \citep{lebouteiller_topological_2022}, which uses Monte-Carlo Markov-Chain (MCMC) techniques \modif{from the python package PyMC \citep{salvatier_pymc3_2016}} with a sampler (sequential Monte-Carlo) adapted to multi-modal posterior distribution. We use the same distribution for the primary parameters as in \citet{ramambason_modeling_2024} for consistency purpose (see Table\,\ref{tab:SFGX_sum}). The density ($n$) and ionization parameter ($U$) therefore follow power-law distributions, while the cut parameter follows a broken power-law with a pivot fixed at the ionization front (i.e., cut=1). The other parameters are described by a single value, with $L_{\rm X}$ and $T_{\rm X}$ being linearly interpolated, while the metallicity, stellar age and total luminosity use a nearest neighbour interpolation. \modiftwo{We obtain a fairly good agreement between the predicted line fluxes and the observations, with 87\% of the line fluxes being reproduced within 3$\sigma$ (see Appendix\,\ref{sec:Ap_linescomp}).}

We obtain probability density functions (PDFs) for each parameter in the model. We also have access to the PDFs of the photoelectric effect, photoionization of hydrogen and helium (from UV and X-ray photons), and CR heating rates. We verified that they account together for at least $90$\% of the total heating in the neutral medium.

\begin{figure}
        \centering
        \includegraphics[width=.5\textwidth]{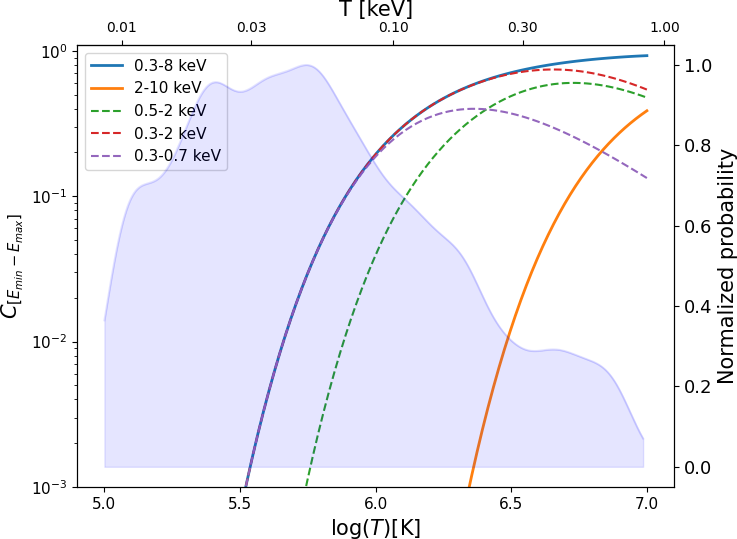}
        \caption{Correction factor for different energy bands to obtain the bolometric X-ray luminosity, depending on the temperature of the inner side of the accretion disk as defined in \citet{mitsuda_energy_1984}. The inferred PDF of $T_{\rm X}$ for the DGS is shown in blue.}
        \label{fig:correction_factor}
    \end{figure}

The inferred values of $T_{\rm X}$ and $L_{\rm X}$ allow us to compute the flux of the X-ray sources, and then to compare it to the observed fluxes when available. However, the comparison is not straightforward, since we infer the bolometric flux of the X-ray source while observations correspond to a given band. We calculate the fraction of the bolometric flux in any given band \modiftwo{for an accretion disk}, $C_{[E_{\rm min}, E_{\rm max}]}$, that depends on the band $[E_{\rm min}, E_{\rm max}]$, and on the inner temperature of the disk $T_{\rm X}$. We use the formula for the accretion disk spectrum in \citet{mitsuda_energy_1984}, in order to compute this correction factor, which is then defined as follows:
\begin{equation}
    { C_{[E_{\rm min}, E_{\rm max}]}(T_{X}) = \dfrac{L_{[E_{\rm min}, E_{\rm max}]}}{L_{\rm tot}} = \dfrac{ \int\limits_{E_{\rm min}}^{E_{\rm max}} \mathrm{dE} \int\limits_{T_{\rm out}}^{T_{X}} \left(\dfrac{T}{T_{\rm X}}\right)^{-11/3} B(E,T) \dfrac{\mathrm{dT}}{T_{\rm X}}  }{\int\limits_0^\infty \mathrm{dE} \int\limits_{T_{\rm out}}^{T_{\rm X}} \left(\dfrac{T}{T_{\rm X}}\right)^{-11/3} B(E,T) \dfrac{\mathrm{dT}}{T_{\rm X}}} }, \label{eq:corr_fact}\end{equation}
where $B(E,T)$ is the blackbody radiation for a temperature $T$, while $T_{\rm X}$ and $T_{\rm out}$ are defined in Sect.\,\ref{subsec:sfgx}. 
It is shown for different bands in Figure\,\ref{fig:correction_factor}.

\section{Results}\label{sec:results}

\subsection{Ionized gas heating processes}\label{subsec:H+heating}

In the ionized gas, we expect the main heating mechanism to be photoionization by UV photons (as illustrated for a typical 1D model in Fig.\,\ref{fig:heatcool1D}). This is confirmed by our models, which show that photoionization accounts for $85$\% of the total heating at high metallicity and up to $\approx100$\% at low metallicity. The remaining heating is due to the photoelectric effect on grains.

\subsection{Neutral gas heating processes}\label{subsec:heating}

\begin{figure}
        \centering
        \includegraphics[width=.5\textwidth]{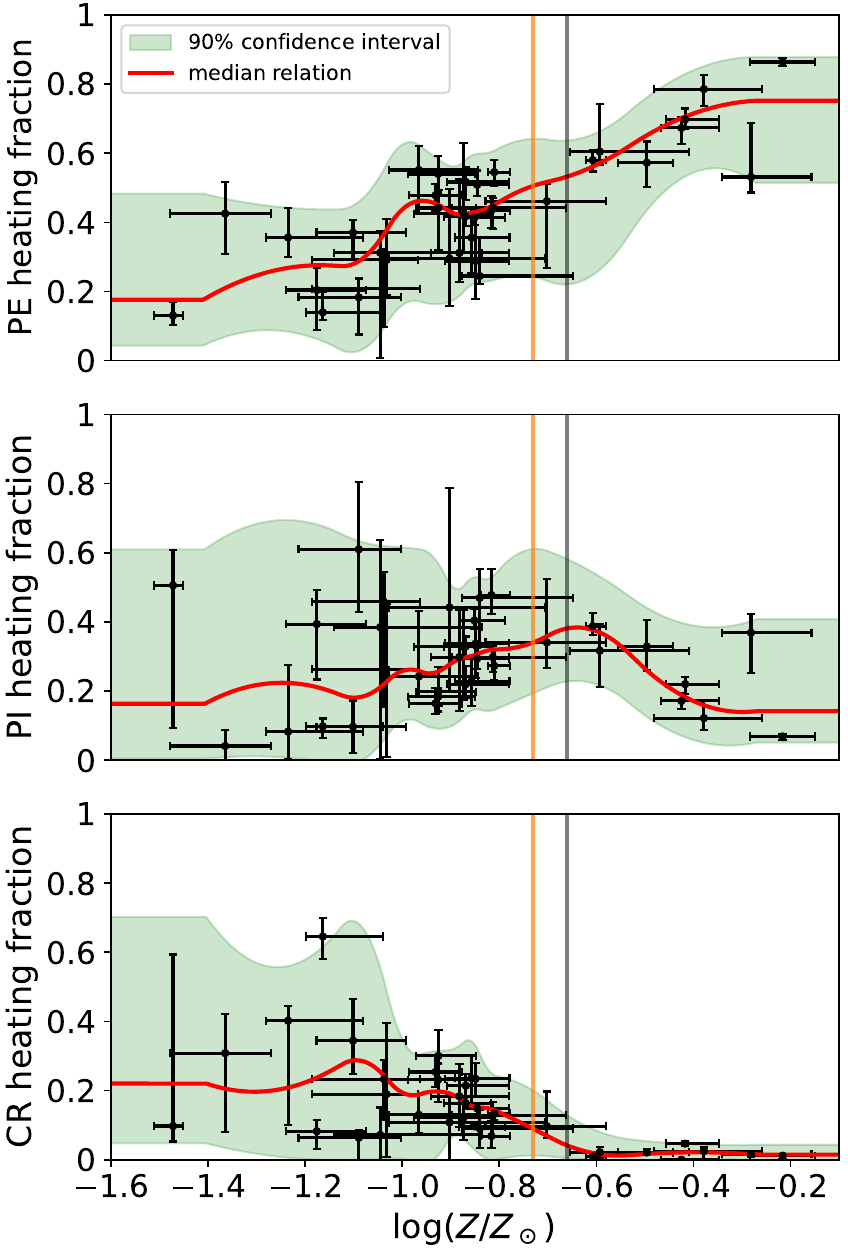}
        \caption{Relative importance of each heating mechanism in the neutral medium with respect to metallicity, for the 37 galaxies of our sample. Top: Photoelectric effect heating. Middle: Photoionization heating. Bottom: Cosmic ray heating. The green shaded area indicates the $90$\% confidence interval computed on the whole sample, with the median relation plotted in red. The black vertical line is the limit above which photoelectric heating dominates ($\approx 22\%$~\Zsol). No PAHs are currently detected in the DGS below the orange vertical line. }
        \label{fig:heating_vs_Z}
    \end{figure}

Even though our models account for molecular gas, we limit our study to neutral atomic gas due to lack of molecular tracers.
In the neutral atomic gas, we compute for each galaxy the heating fraction due to the photoelectric effect, photoionization, and cosmic rays, defined as the heating rate of one mechanism integrated over the whole H\,{\sc i} region \modif{(between cut parameter values 1 and 2, as defined in Sect.\,\ref{subsec:sfgx})} divided by the total heating rate in the H\,{\sc i} region. These heating fractions are shown in Figure\,\ref{fig:heating_vs_Z}.

We find that the photoelectric effect heating fraction increases with metallicity and becomes predominant at metallicity above $\approx 22\%$\,\Zsol. Interestingly, this metallicity lies close to the detection limit of PAHs in the DGS, at a metallicity of $\approx 19\%$\,\Zsol. The photoionization heating fraction shows no clear trend with metallicity, but remains significant at all metallicities, being even above $50$\% for some galaxies. Cosmic rays only heat significantly at metallicities below $1/6$\,Z$_\odot$, while accounting for only a few percent of the heating at high metallicity.

\subsection{Inferred X-ray luminosities}\label{subsec:Xluminosity}

\begin{figure}
    \centering
    \includegraphics[width=.5\textwidth]{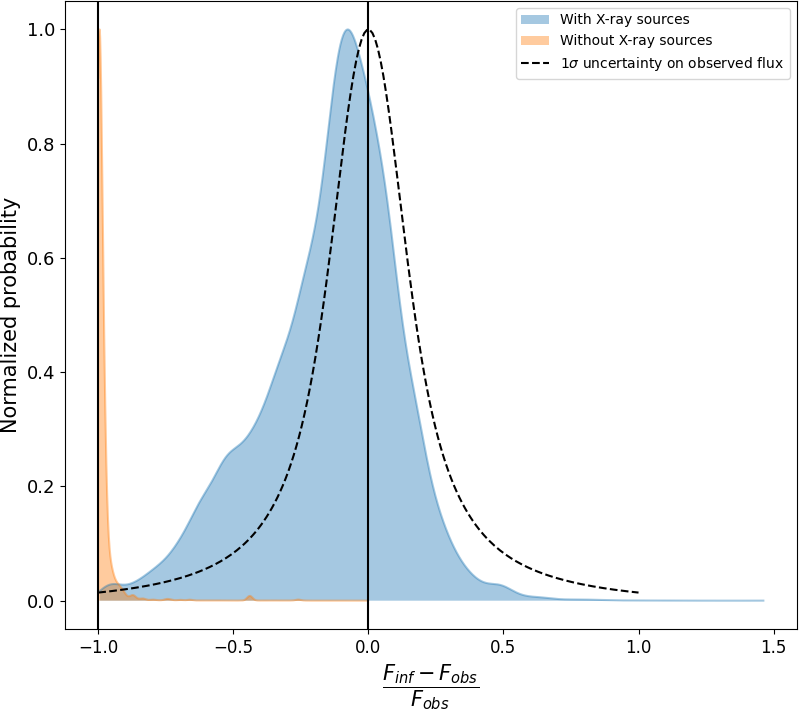}
    \caption{\OIV\ inferred flux vs. observed flux for the DGS, with (blue) and without (orange) X-rays sources. A value of $0$ means that the flux is perfectly reproduced, $-1$ means that no flux is predicted. The dashed line shows the uncertainty distribution on the measured flux.}
    \label{fig:OIV_X}
\end{figure}

Figure\,\ref{fig:OIV_X} shows that X-ray sources are necessary to reproduce the observed \OIV\ flux. \modif{Indeed, when removing the X-ray sources from the models, no significant \OIV\ flux is predicted. When X-ray sources are present however, the predicted flux is compatible with the observations within $1\sigma$ for most of the galaxies.} BPASS stellar SEDs cannot account for this emission, even when taking Wolf-Rayet stars into account. However, this relies on the BPASS stellar model hypotheses, and other results could be achieved using other prescriptions (see, e.g., \citealt{lecroq_nebular_2023}). 

\begin{figure}
    \centering        
    \includegraphics[width=.48\textwidth]{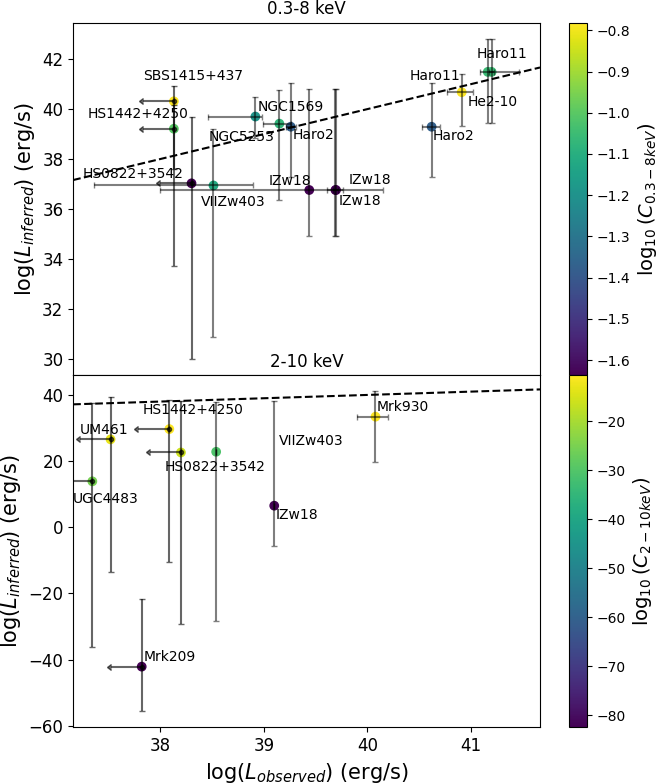}
    \caption{Inferred vs.\ observed X-ray luminosity. Top: $0.3-8$\,keV. Bottom: $2-10$\,keV. The dashed line indicates the 1:1 ratio. \modif{The colorscale shows the logarithm of the correction factor used to calculate the inferred luminosity, as defined in Equation\,\ref{eq:corr_fact} and shown in Figure\,\ref{fig:correction_factor}.}}
    \label{fig:infered_X_lum}
\end{figure}

The question is then whether the inferred X-ray fluxes in our models are compatible with the observations. We focus only on two bands often used in the literature, $0.3-8$\,keV (low energy) and $2-10$\,keV (high energy) since ULXs have already been detected in these two bands in some galaxies of our sample (see Section\,\ref{subsec:Xobs}). 
We see in Figure\,\ref{fig:infered_X_lum} that, despite large uncertainties, all the predicted fluxes in the $0.3-8$~keV band are compatible with the observations, even within $1$ dex for 5 galaxies. We verified that the predicted fluxes are not driven by the edges of the grid. In the $2-10$~keV band, however, the predicted fluxes are heavily underestimated. We discuss further these results in Section\,\ref{subsec:Xsources}.

\section{Discussion}
\label{sec:discussion}

\subsection{Photoelectric effect}
\label{subsec:PEheating}

The increasing trend of the photoelectric heating fraction with metallicity is somewhat expected, since the DGR is known to increase with metallicity and our models do force it to follow the median relation from \citet{galliano_nearby_2021}. However, this prescription being constrained from other observations (in particular mid- and far-IR photometry), we expect the photoelectric effect heating fraction not to suffer from large systematic uncertainties. In order to ensure that our results are not biased by \modif{the fixed DGR-Z relationship}, we ran a subsample of $20$ galaxies \modif{for which \citet{galliano_nearby_2021} measured the DGR}. Despite small individual variations, all $20$ galaxies fall within the $90$\% confidence envelop that we computed from the previous runs where the DGR was fixed (see Fig\,\ref{fig:heating_vs_Z}). This holds for the photoelectric effect, photoionization, and cosmic ray heating fractions, showing that our results are robust to a change in the DGR.

\begin{figure}
    \centering
    \includegraphics[width=.45\textwidth]{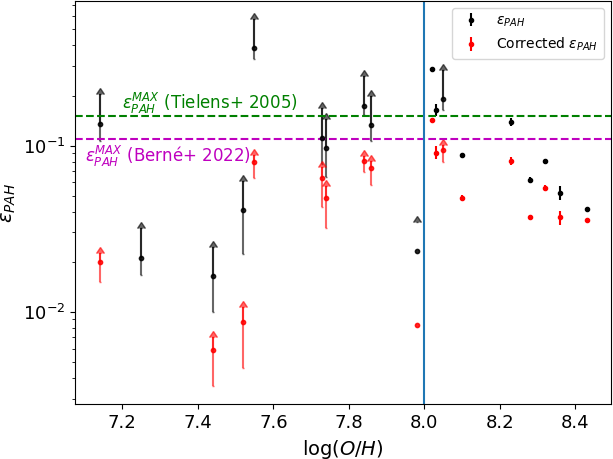}
    \caption{PAH heating efficiency for galaxies in the subsample of the DGS with PAH observations. The observed $\epsilon_{\rm PAH}$ are shown with black points, the corrected $\epsilon_{\rm PAH, corrected}$ values are shown in red. The theoretical maximal value of $15$\% is indicated by the green line and the maximum observed value in \citet{berne_contribution_2022} in magenta. The vertical blue line indicates the current limit below which PAH are undetected in the DGS.}
    \label{fig:Epah}
\end{figure}

One possible application of our work is to probe the photoelectric effect heating efficiency on PAHs, which is the fraction of the energy received by the PAHs that goes into ISM heating. An often used observational proxy (e.g., \citealt{okada_probing_2013, berne_contribution_2022}) is 
\begin{equation}
\epsilon_{\rm PAH} = \dfrac{[{\rm CII}]~158~\mu m+[{\rm OI}]~63~\mu m}{L_{\rm PAH} + [{\rm CII}]~158~\mu m + [{\rm OI}]~63~\mu m},\label{eq:Epah}\end{equation} 
with reported values below $15$\%, which coincides with the theoretical maximum value calculated in \citet{tielens_physics_2005}. However, when applied to low-metallicity galaxies in our sample, values above $15$\% and even above $38$\% for one galaxy are observed (see Fig.\,\ref{fig:Epah}), hinting that $\epsilon_{\rm PAH}$ is not an accurate proxy for the photoelectric effect heating efficiency at low metallicity. This can be explained by the fact that the sum \CII\ + \OI\ traces the total \modif{cooling of the gas which may be heated through various mechanisms. Since $\epsilon_{\rm PAH}$ is therefore overestimated due to the assumption of a dominant photoelectric effect heating, we may try to use the photoelectric effect heating fraction inferred from our modelling, in order to derive a corrected value:} \begin{equation}
\epsilon_{\rm PAH, corrected} = \dfrac{\alpha([{\rm CII}]+[{\rm OI}])}{L_{\rm PAH} + \alpha([{\rm CII}] + [{\rm OI}])}\label{eq:Epah_corr}\end{equation} with $\alpha$ the fraction of heating due to the photoelectric effect, computed using MULTIGRIS predictions (Fig.\,\ref{fig:heating_vs_Z}, top panel).
The values of $\epsilon$ with and without correction are shown in Figure\,\ref{fig:Epah}. The $5$ galaxies with $\epsilon_{\rm PAH}>15\%$ have $\epsilon_{\rm PAH, corrected}$ values dropping below this threshold.
We note that while our correction always provides values below $15$\%, it is biased by the fact that the photoelectric heating fraction in our models accounts for PAH heating but also heating on small grains that are not PAHs. While this last component is likely negligible at high metallicity \citep{berne_contribution_2022}, it might not be the case anymore at significantly low metallicity. There are indeed clues that smaller, non-PAH, grains are present in low-metallicity galaxies, due to the fragmentation of big grains by supernovae \citep{galliano_interstellar_2018}.
We cannot exclude that such grains contribute to the photoelectric effect heating, effectively lowering the photoelectric effect efficiency, but our specific study cannot conclude on this.

\subsection{Cosmic rays}\label{subsec:CR}

\begin{figure}[H]
    \centering
    \includegraphics[width=.48\textwidth]{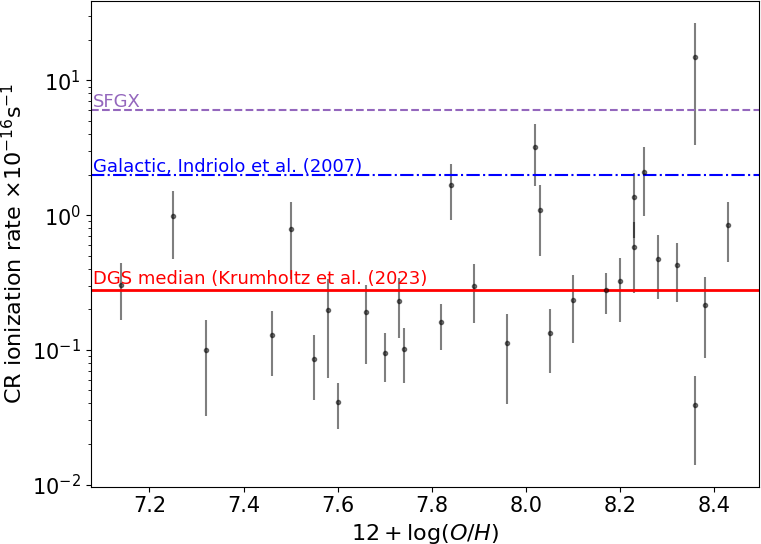}
    \caption{Cosmic ray ionization rate vs. $Z$ in the DGS, following the prescription from \citet{krumholz_cosmic_2023}. The solid line is the median CR ionization rate, the dashed line the CR ionization rate used in SFGX, and the dotted line the galactic CR ionization rate from \citet{indriolo_documentclassaastex_2007}.}
    \label{fig:CRIR}
\end{figure}

One of the main limitations of the present study is the assumption of a single, fixed, cosmic ray ionization rate in SFGX ($6\times 10^{-16}\,\mathrm{s}^{-1}$; see Section\,\ref{subsec:sfgx}). The CR heating fraction is therefore driven by the density of the medium and the depth within the cloud ("cut" parameter). We consider that this CR ionization rate is in fact probably overestimated. Following the prescription from \citet{krumholz_cosmic_2023} ($\zeta_{\rm tot} = 10^{-16} \left(\dfrac{\rm M_{gas}/SFR}{\rm Gyr}\right)^{-1} \mathrm{s^{-1}}$ with $\zeta_{\rm tot}$ the CR ionization rate) based on the SFR and the mass of neutral gas (H$^0$+H$_2$ measured in \citealt{remy-ruyer_gas--dust_2014, remy-ruyer_linking_2015}), we find a median value for the CR ionization rate in the DGS of $3\times 10^{-17}\,\mathrm{s}^{-1}$ (see Fig.\,\ref{fig:CRIR}). \modiftwo{Similar values are found by Brugaletta et al. (arXiv:2410.19087) in magneto-hydrodynamical simulations of low-metallicity galaxies.}
\modiftwo{Since there is a competition between the ionization from X-rays and CR in the neutral gas, overestimating the CR heating leads to underestimating the X-ray heating. Although we cannot provide a quantitative value for how much the X-ray heating is actually underestimated, this strenghtens the fact that X-rays are a significant, or even dominant source of heating in the neutral gas at low metallicity.}

\modif{We ran Cloudy models with the CR ionization rate decreased by a factor of $30$, which corresponds to the difference between the SFGX CR ionization rate and the DGS median CR ionization rate, the latter calculated from \citet{krumholz_cosmic_2023} using the SFR and $M_{\rm gas}$ from \citet{remy-ruyer_gas--dust_2014}. The other primary parameters are chosen to match the median values inferred for the DGS. Performing this calculation for $6$ different metallicities, we find that the CR heating fraction decreases by a factor of about $20$ for very low metallicity models ($1/20$\,Z$_\odot$) up to $60$ for medium metallicity models ($1/5$\,Z$_\odot$). For metallicities above $1/3$\,Z$_\odot$, the CR heating fraction is negligible when using the low CR ionization rate. For the full metallicity range, the photoelectric effect and photoionization heating fraction do not change significantly between the high CR ionization rate and low CR ionization rate models, showing the same trend with metallicity as in Figure\,\ref{fig:heating_vs_Z}.}

The extinction depth of the $\mathrm{H^+}$-$\mathrm{H^0}$ and $\mathrm{H^0}$-$\mathrm{H_2}$ transitions does not change significantly between the models with the two CR ionization rates, which implies that the integration over the whole H{\sc i} region corresponds to the same physical volume, and also implies that the test results are indeed meaningful. \modif{However, one should keep in mind that changing the CR ionization rate in SFGX is likely to modify the final solution, making it difficult to predict the exact impact of such a change on the final heating fraction. For this test, we verified the impact of decreasing the CR ionization rate on the intensity of the lines used to constrain the models. While no significant change is observed for metallicities above $1/10$\,Z$_\odot$ , some line intensities do change significantly for lower metallicity models, implying that the inference should eventually be performed again with an extended grid.}

\begin{figure}
    \centering
    \includegraphics[width=.45\textwidth]{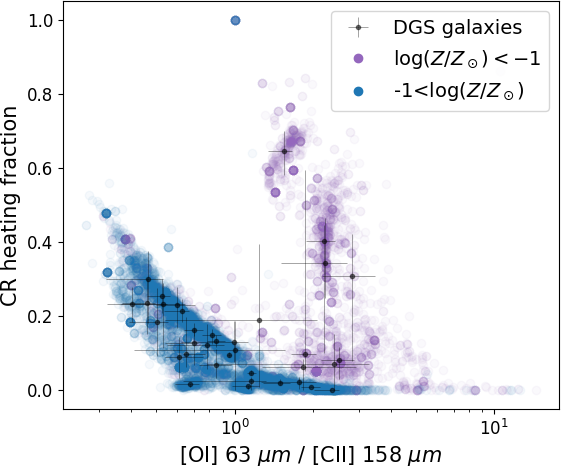}
    \includegraphics[width=.45\textwidth]{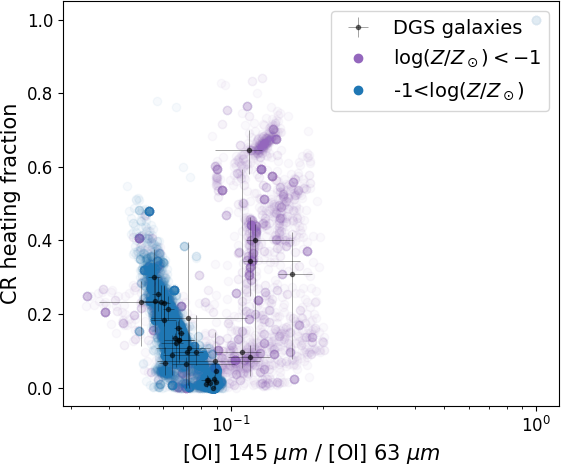}
    \caption{CR heating fraction vs. \OI / \CII\ (top) and \OIb / \OI\ (bottom). Each point represents a model randomly drawn from the inferred posterior distribution. The global values for each galaxy are overplotted in black. }
    \label{fig:CRlines}
\end{figure}

Even considering these caveats, we wish to understand how the cosmic ray heating fraction is constrained. We find that some line ratios correlate better with the CR heating fraction, especially \CII/\OI\ and \OIb/\OI, as shown in Figure\,\ref{fig:CRlines} for the moderately \modif{metal-poor galaxies ($>1/10$\,Z$_\odot$)}, which is the range of metallicities where most of the galaxies in our sample lie. Even though the CR heating fraction is linked to the photoelectric effect and photoionization heating fraction, there is no correlation between these line ratios and the photoelectric effect nor the photoionization heating fraction. Moreover, these correlations hold even if X-ray sources are removed from the models, and the correlations are observed at every depth of the cloud in the \modiftwo{atomic gas}. \modif{However, the fact that our grid considers a single CR ionization rate -- that may be overestimated -- somewhat biases these correlations, making it difficult to draw a robust conclusion.} We have not identified clear correlations between the photoionization heating fraction and any specific line ratio. For lower metallicities, no correlation is seen between the cosmic ray heating fraction and line ratios.

\subsection{Photoionization}\label{subsec:PIHeating}

The main caveat regarding the photoionization heating fraction is that we cannot differentiate UV photons (primarily coming from the stellar population continuum) from X-ray photons (primarily coming from X-ray sources associated with the multi-color blackbody prescription), since Cloudy does not track down photons emitted by any given continuum source. Therefore, we are not able to determine the exact, specific, contribution of X-ray sources to neutral medium heating. 
To overcome this limitation we focus on the best models inferred with MULTIGRIS either with and without X-ray sources and compare them. While we are unable to obtain a quantitative value for the heating rate due to X-ray photons specifically (compared to UV photons), these tests show that \modif{all the galaxies are compatible with a higher photoionization heating fraction when X-ray sources are considered}, with a median ratio (photoionization heating fraction with X-ray sources)/(photoionization heating fraction without X-ray sources) of $1.7$. The results imply that X-rays contribute at least significantly to the neutral medium heating.

\subsection{X-ray source properties}\label{subsec:Xsources}

Despite an overall agreement between our predicted fluxes and the observations in the $0.3-8$~keV band, the predicted values suffer from large uncertainties. These uncertainties come from the two parameters used to derived the flux, namely $L_{\rm X}$ and the correction factor $C_{[0.3-8\,{\rm keV}]}$ (defined in Section\,\ref{subsec:inference}), which depends on the temperature $T_{\rm X}$. While $L_{\rm X}$ is well constrained in most of the galaxies, $T_{\rm X}$ is much more uncertain. Moreover, $C_{[0.3-8\,{\rm keV}]}$ strongly varies with the temperature, especially at low $T_{\rm X}$ ($<5.8$ in log), which is the case for most ($22$) galaxies in our sample. In this temperature range, a rather small uncertainty on $T_{\rm X}$ can lead to a large uncertainty on the predicted flux, making $T_{\rm X}$ the main source of uncertainty in the predicted fluxes. 

The fact that $T_{\rm X}$ is not well constrained is mainly due to the lack of tracers with high ionization potentials. The highest ionization potential tracer in our sample is O$^{3+}$ ($55$\,eV), with additional constraints from Ne$^{4+}$ ($97$\,eV) via the upper limits on \NeV. This means that we effectively only constrain the low-energy part of the X-ray SED. In our case, however, the X-ray SED is defined by a single source with parameters $L_{\rm X}$ and $T_{\rm X}$, and we are able to constrain the entire SED simply based on the low-energy part. Nevertheless, we would need high enough energy tracers to provide unambiguous and reliable constraints on $T_{\rm X}$: to constrain high $T_{\rm X}$ values above $10^6~\mathrm{K}$, we would need tracers with significantly higher ionization potentials ($\gtrsim100$\,eV). The use of IR coronal lines such as [Fe{\sc vi}] ($75$\,eV) or [Si\,{\sc x}] ($351$\,eV), observable with JWST, have been proposed by \citet{cann_hunt_2018} to constrain the X-ray SEDs of AGN, and they could be relevant in our study. SDSS observations of optical lines such as [Fe{\sc vii}] ($99$\,eV) or [Fe{\sc xi}] ($262$\,eV) \citep{reefe_class_2022} could also prove useful. \modif{Unfortunately, no measurements of these lines are available for the DGS}. An alternative to using high ionization potential species could be to use species around the ionization front or from the neutral gas instead \citep{richardson_optical_2022}.

Another source of uncertainty in the X-ray flux comparison comes from the X-ray source prescription used in SFGX. While we use an accretion disk \citep{mitsuda_energy_1984} -- often used to represent the low-energy ($<2$\,keV) spectrum of ULXs -- such objects are known to display a power-law component at higher energy \citep{gladstone_ultraluminous_2009, pintore_ultraluminous_2014}. This high energy component is not taken into account in SFGX, which explains why we underestimate the predicted $2-10$\,keV fluxes \modif{(see Fig.\,\ref{fig:infered_X_lum})}. New X-ray source models should eventually be implemented in SFGX to account for both the high-energy and low-energy components of the X-ray source SEDs, allowing us to constrain both the $0.3-8$\,keV and $2-10$\,keV band fluxes. It is worth noticing that even if the high-energy component is dominant within the $2-8$\,keV range, it is roughly equal to the accretion disk component within the whole $0.3-8$\,keV range, making our comparison in this band valid.

We emphasize that the predicted $T_{\rm X}$ is not necessarilly actual physical temperature but rather a parameter of our model to describe the X-ray source SED. In particular, it has been shown that using physical models instead of a power-law for the comptonization component can lead to significant differences in $T_{\rm X}$ \citep{gladstone_ultraluminous_2009}. In order to infer the actual physical properties of the source, we would need such a physical model, flexible enough to represent various type of objects (stellar mass black holes, intermediate-mass black holes, AGN), such as models described in \citet{garofali_modeling_2024} for ULXs. 

It is also important to note that even with a physical model covering the whole $0.3-10$\,keV range and enough tracers to constrain it, we do not expect a 1:1 relation between the predicted luminosity and the observations. Indeed, we do not infer the intrinsic X-ray luminosity of the source, but rather the X-ray luminosity seen on average by the surrounding ISM over a cooling timescale. Line-of-sight effects and time variability of the source are then expected to produce a scatter in the relation.

\section{Conclusion}
\label{sec:conclusion}
This paper presents an application of the statistical code MULTIGRIS to a sample of low-metallicity galaxies (DGS; \citealt{madden_overview_2013}) to study the contribution of the photoelectric effect, photoionization by UV and X-ray photons, as well as ionization by cosmic rays to the heating of the neutral atomic ISM. We also attempt to predict the intrinsic X-ray flux for galaxies with known X-ray sources. Our main results are:
\begin{itemize}
    \item The contribution of photoelectric effect heating to the total heating of the neutral gas increases with metallicity, and dominates at high metallicity ($>1/8$\,Z$_\odot$).
    \item Photoionization heating contributes significantly at all metallicities, with a large scatter between individual galaxies. In some cases, it can become the dominant source of heating. 
    \item Cosmic ray heating is negligible at high metallicity. At lower metallicity, it becomes an important mechanism but is usually not the dominant one. However, the exact contribution of cosmic rays remains uncertain, since the cosmic ray ionization rate is forced to a single value in our models. An extension of SFGX with various cosmic ray ionization rates would be necessary to better constrain the impact of cosmic rays on the neutral medium heating, but is beyond the scope of this paper and intended for future studies.
    \item \modiftwo{The single CR ionization rate used in our study is likely overestimated. This implies that the X-ray heating is likely underestimated, which is especially important for low metallicity galaxies where the X-ray heating may in fact dominate the total heating.}
    \item We find that the heating efficiency of the photoelectric effect on PAHs measured from observational proxies is higher than the theoretical maximal value using the $\epsilon_{\rm PAH}$ proxy. However, it becomes compatible for all our galaxies when accounting for the inferred fraction of photoelectric effect heating. This implies that the usual proxy for the PAH heating efficiency must be taken with caution in low-metallicity galaxies. 
    \item The luminosity of X-ray sources can be inferred from IR lines of species with high ionization potentials (H{\sc ii} region) and/or of species in the neutral gas. We are able to recover reasonably well the observed \modif{$0.3-8$\,keV} fluxes of ULXs in the galaxies of the DGS despite large uncertainties. We determine that uncertainties are dominated by the temperature of the accretion disk and propose potential tracers to better constrain our predicted X-ray fluxes.
    \item A correlation is found between the cosmic ray heating fraction and neutral gas line ratios involving \OI, \OIb, and \CII\ for galaxies with metallicity between $0.08$ and $0.3$\,Z$_\odot$. However, this correlation is most likely biased by the fact that we use a single cosmic ray ionization rate, with a value probably overestimated in our grid of models. 
\end{itemize}

The method used in this work offers an interesting perspective to study the gas heating in other types of galaxies, especially at high redshift since we aim to disentangle the different ISM phases from an integrated spectrum. Globally, this method is well adapted to high-$z$ observations with ALMA, which include neutral gas tracers such as \CII\ and \OI\, as well as potential future far-IR spectroscopic facilities such as PRIMA. Our study suggests that X-ray sources are important contributors to the heating of the star forming gas reservoir and could be useful ingredients to implement in galaxy simulations. 

The same method can also be applied to study specifically the X-ray source properties present in low-metallicity galaxies. In particular, it offers a potential way to probe the X-ray spectrum shape by probing only the surrounding ISM emission, to characterize X-ray sources without a direct X-ray detection.

\begin{acknowledgements} LR gratefully acknowledges funding from the Deutsche Forschungsgemeinschaft (DFG, German Research Foundation) through an Emmy Noether Research Group (grant number CH2137/1-1). CR gratefully acknowledges the support of the Elon University Japheth E. Rawls Professorship. This work is supported by the FACE Foundation Transatlantic Research Partnership Fund (award TJF21\_053).
\end{acknowledgements}

\bibliographystyle{aa}
\bibliography{MyLibrary}

\begin{appendix}
\section{Agreement between models and observations}
\label{sec:Ap_linescomp}

\begin{figure}[h]
    \centering
    \includegraphics[width=2\linewidth]{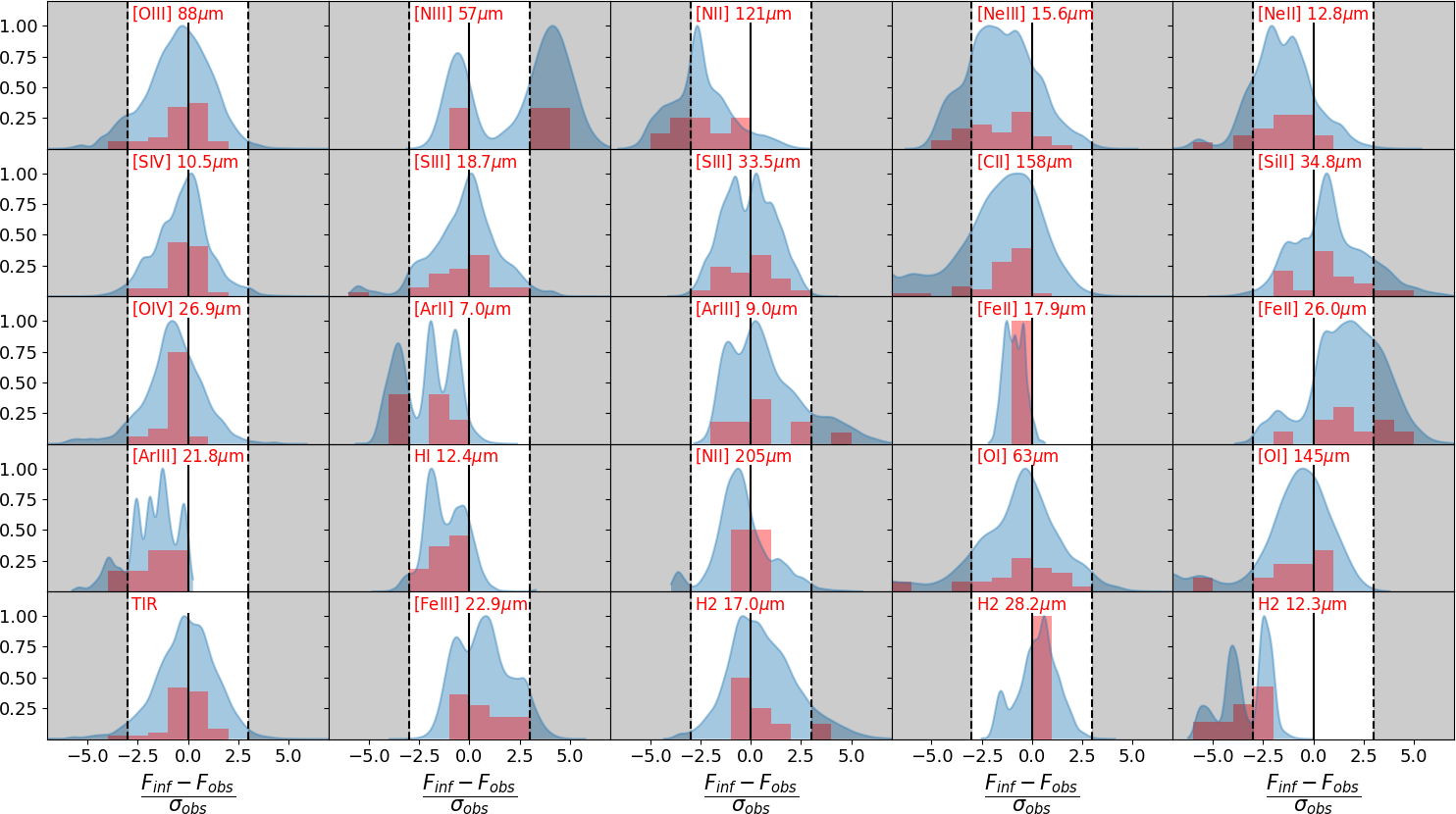}
    \caption{Probability density function of the difference between observed and predicted lines for the whole sample. The shaded grey area shows the predicted points that fall further than $\pm 3 \sigma$ from the observation. The blue PDFs shows the concatenated sample while the red histograms shows the distribution of the median flux predicted for each galaxy.}
    \label{fig:line_comparison}
\end{figure}

In order to caracterize how well our models reproduce the observations, we computed for each observed line in the sample the difference between the observed flux and the predicted flux as $\Delta = \dfrac{F_{\rm inferred}-F_{\rm observed}}{\sigma_{\rm observed}}$ (see Fig.\,\ref{fig:line_comparison}). We find that for most of the lines, the predicted flux fall within $3\sigma$ of the observation for almost all the galaxies. When plotting the PDF of $\Delta$ for the whole sample, we find that it peaks is within the $3 \sigma$ area for all the lines except [N{\sc iii}] $57$\,\mum\ and H$_2$ $12.3$\,\mum. Over the whole sample, $87\%$ of the line fluxes are reproduced at the $3\sigma$ level and $55\%$ at the $1 \sigma$ level.

\end{appendix}

\end{document}